\documentclass[review]{elsarticle}

\usepackage{lineno,hyperref}
\usepackage{amsmath} 
\usepackage{amssymb} 
\modulolinenumbers[5]

\journal{Physics Letters A}

\begin{document}

\begin{frontmatter}

\title{Explicit, analytical radio-frequency heating formulas for 
spherically symmetric nonneutral plasmas in a Paul trap 
}

 
 

 
\author[A1,A2,A3]{Y. S. Nam\fnref{myfootnote}} 
\fntext[myfootnote]{present location at IonQ Inc., 4505 Campus Drive, College Park, MD 20740, USA}
\author[A1,A4]{D. K. Weiss} 
\author[A1]{R. Bl\"umel\corref{mycorrespondingauthor}} 
\cortext[mycorrespondingauthor]{Corresponding Author:}
\ead{rblumel@wesleyan.edu}
 
\address[A1]{Department of Physics, Wesleyan University, Middletown, CT 06459, USA}
\address[A2]{Institute for Advanced Computer Studies,   
University of Maryland, College Park, MD 20740} 
\address[A3] {Joint Center for Quantum Information and Computer Science, 
3100 Atlantic Building, University of Maryland, College Park, MD 20742}
\address[A4]{Department of Physics and Astronomy, Northwestern University, 
Evanston, IL 60208, USA}

\begin{abstract}
We present explicit, analytical heating formulas that predict 
the heating rates of spherical, nonneutral plasmas stored in a Paul trap 
as a function of cloud size $S$, particle number $N$, and 
Paul-trap control parameter $q$ in the low-temperature 
regime close to the cloud $\rightarrow$ crystal 
phase transition. We find excellent agreement 
between our analytical heating formulas and detailed, time-dependent 
molecular-dynamics simulations of the trapped plasmas. 
We also present the results of our numerical solutions of a 
temperature-dependent mean-field equation, which are 
consistent with our 
numerical simulations and our analytical results. This is the 
first time that analytical heating formulas are presented that 
predict heating rates with reasonable 
accuracy, uniformly for all $S$, $N$, and $q$. 
\end{abstract}


\end{frontmatter}

\linenumbers

\section{Introduction}
\label{INTRO}
The Paul trap \cite{Paul1,Ghosh} has long secured its place as 
an indispensable tool in many fields of science 
with applications ranging from 
atomic clocks \cite{atom_clock} and quantum  
computers \cite{qc1,qc2,qc3} to mass 
spectrometry \cite{chem} and 
particle physics \cite{std_model}. Given its long 
and successful history, it is surprising that the Paul trap 
still offers unsolved, fundamental physics problems. 
For instance, if $N\geq 2$ charged particles are 
simultaneously stored in a Paul trap, the 
kinetic energy of the 
particles increases in time, i.e., 
they exhibit the phenomenon of 
radio-frequency (rf) heating. Rf heating cannot be 
switched off. 
It is a basic physical process 
that necessarily accompanies the 
operation of the trap.  
Surprisingly, although the existence of rf 
heating has been known \cite{rf1,rf2,rf3} 
and studied \cite{rf4,BKQW,Tarnas,Jones} for a long time, 
explicit heating 
formulas, capable of predicting 
rf heating rates, have so far not been available. 
This paper addresses this deficiency. 
In particular, focusing on trapped 
clouds of charged particles of the 
same sign of charge (known as one-component, 
nonneutral plasmas \cite{Davidson}), we present 
explicit, 
analytical heating formulas that predict the 
heating rates of spherical, one-component, 
nonneutral plasmas consisting of $N$ charged particles 
as a function of cloud size $S$ and Paul-trap 
parameter $q$ \cite{Paul1,Ghosh,BKQW}.  
 
We define rf heating as the cycle-averaged 
power extracted from the rf field of the trap. 
There are two situations 
of theoretical and experimental interest. 
(A) With the help of a cooling mechanism, such as 
buffer-gas cooling \cite{rf1,buffer} or 
laser cooling \cite{rf4,BKQW}, 
the plasma may be brought to a stationary state 
in which the size of the ion cloud stays constant, 
on average, over extended periods of time. 
(B) After the stationary state is reached, 
the cooling may be switched off. From this point on, 
due to the nonlinear nature of the particle-particle 
interactions in the plasma \cite{BKQW}, 
the plasma cloud will absorb energy from the 
rf trapping field, heat up, and expand. 
As discussed in Section~\ref{DISC}, 
the heating rates in situations (A) and (B) 
are different, since in situation (A)  
the rf field has to provide additional power to 
counteract the dissipative losses due 
to the micromotion \cite{Ghosh} of the 
trapped particles. 
To keep the discussion focused, we concentrate 
in this Letter 
on situation (A) and comment briefly on 
situation (B) in Section~\ref{DISC}. 
 
This Letter is organized as follows. In 
Section~\ref{THEORY} we present the 
basic equations that underlie our theory 
of rf heating.   
In Section~\ref{RESULTS} we 
present our 
analytical and numerical methods together 
with a 
detailed comparison between 
numerically and analytically computed  
rf heating rates. Excellent agreement 
between the results of our numerical simulation data 
and our analytical 
rf heating formulas is obtained. 
In Section~\ref{DISC} we discuss our 
results. We conclude our paper in 
Section~\ref{CONC}. For 
the convenience of the reader  
we also provide 
an appendix, in which 
we convert our dimensionless quantities and 
results to standard SI units. 

\section{Theory} 
\label{THEORY} 
 
The starting point of our work is the following set 
of dimensionless equations of motion that 
describe the motion of $N$ charged particles in a hyperbolic 
Paul trap \cite{Jones} 
\begin{equation}
\ddot{\vec r}_i + \gamma \dot{\vec r}_i + 
[a-2q\sin(2 t)]
 \left( \begin{matrix} x_i \\ y_i \\ -2 z_i \end{matrix} \right) 
 =   \sum_{\substack{j=1\\j\neq i}}^N 
 \frac{\vec r_i - \vec r_j}{|\vec r_i - \vec r_j |^3}, \ \ \ i=1,\ldots,N. 
\label{THEO1} 
\end{equation} 
Here ${\vec r}_i=(x_i,y_i,z_i)$ denotes the position vector of particle 
number $i$, $\gamma$ is the damping constant, $t$ is 
the time, and $a$ and $q$ are the two 
dimensionless control parameters of 
the Paul trap \cite{Paul1,Ghosh,BKQW}. 
The solutions ${\vec r}_i(t)$ of 
(\ref{THEO1}) are 
best represented as a superposition of a slow, large-amplitude 
macromotion \cite{Ghosh}, 
$\vec R_i(t)=(X_i(t),Y_i(t),Z_i(t))$, 
and a fast, small-amplitude 
micromotion \cite{Ghosh}, $\vec \xi_i(t)$, i.e., 
\begin{equation}
{\vec r}_i(t) = \vec R_i(t) + \vec \xi_i(t), 
\label{THEO1a}
\end{equation}
where, to lowest order, 
\begin{equation}
\vec \xi_i(t) = -\frac{q}{2} \sin(2t) 
\left( 
\begin{matrix}  X_i(t) \\ 
                Y_i(t) \\
            -2 Z_i(t) \\ 
\end{matrix} 
\right) .
\label{THEO1b}
\end{equation}
The damping constant $\gamma$ in (\ref{THEO1}) 
plays a dual role. In our numerical simulations 
we use $\gamma$ as a convenient way to achieve 
a spherical, stationary state in which rf heating balances 
the cooling induced by $\gamma$.
As shown in 
\cite{Jones}, we may then use the equality between 
heating and cooling in the stationary state to 
compute the rf heating rate according to 
\begin{equation}
H = \frac{dE}{dt} = 2\gamma E_{\rm kin} , 
\label{THEO2} 
\end{equation}
where $E$ is the total cycle-averaged energy of the plasma cloud and 
$E_{\rm kin}$ is its cycle-averaged kinetic energy \cite{Jones}. 
Since in this 
paper we focus on spherical trapped 
plasma clouds, and since spherical 
clouds are obtained for the choice 
$a=q^2/2$ \cite{Ghosh,BKQW}, we assume this 
setting of the parameter $a$ for the remainder 
of this paper. Spherical clouds in the 
stationary state are 
conveniently characterized by their size, 
\begin{equation}
S = \lim_{n\rightarrow\infty} \frac{1}{n} 
\sum_{k=1}^n
\left[\sum_{i=1}^N \vec R_i^2(k\pi)\right]^{1/2}, 
\label{THEO3}
\end{equation}
where $\vec R_i$ is evaluated at multiples 
of $\pi$, where, according to (\ref{THEO1b}), 
the micromotion amplitude vanishes. 
 
\section{Methods and results} 
\label{RESULTS} 
%
\begin{figure}
\includegraphics[scale=0.5,angle=0]{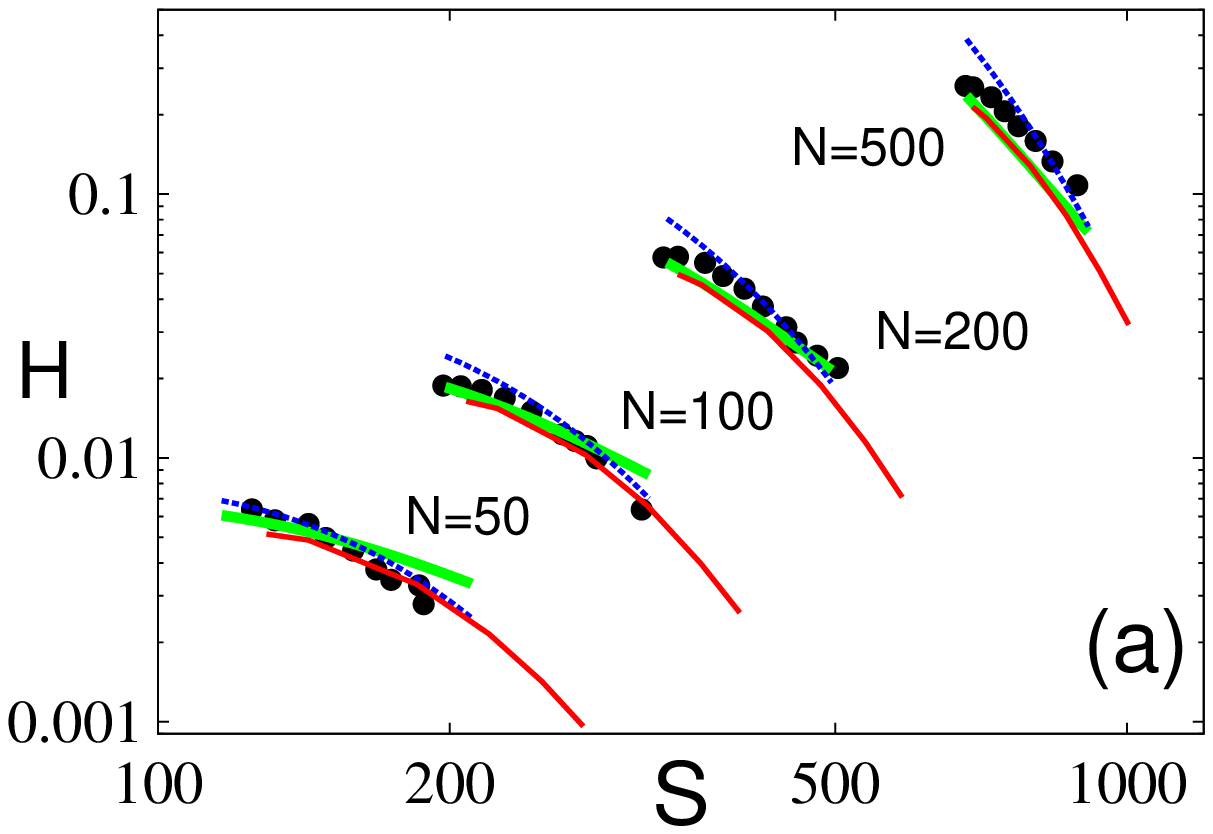}
\includegraphics[scale=0.5,angle=0]{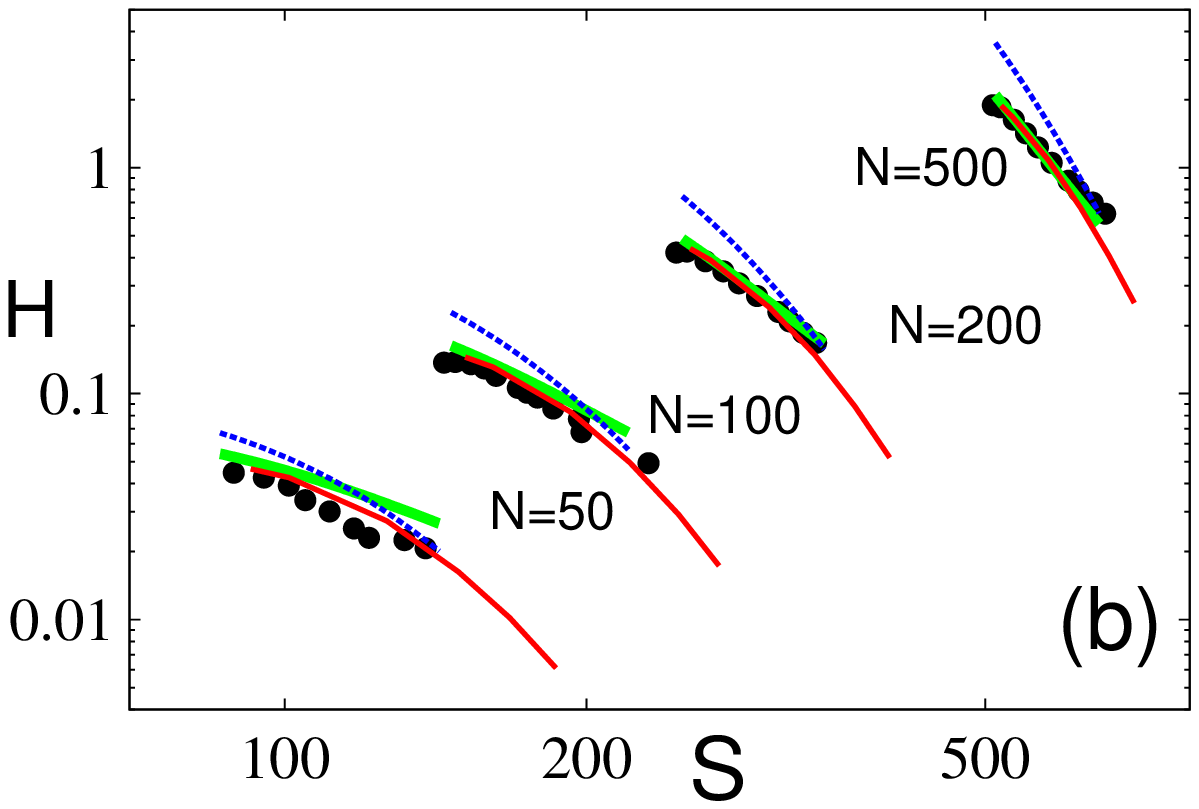}
\includegraphics[scale=0.5,angle=0]{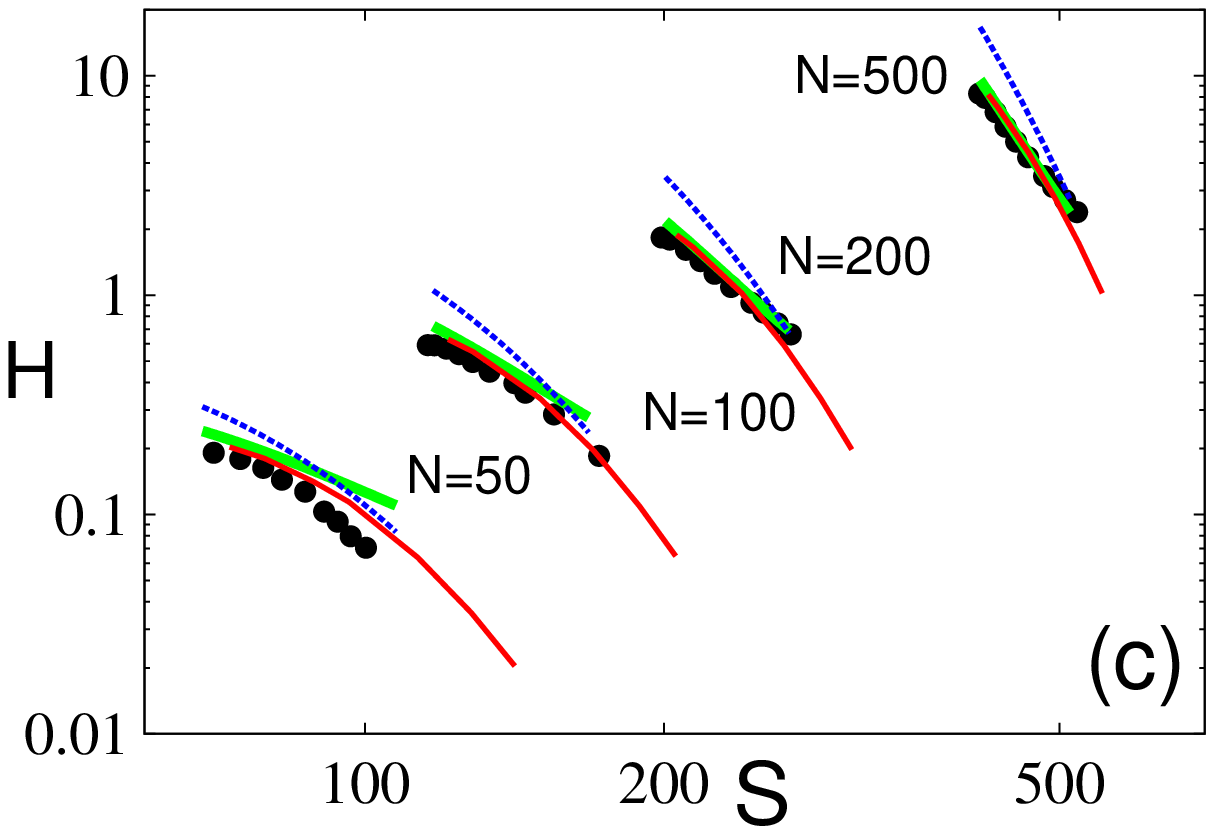}
\includegraphics[scale=0.5,angle=0]{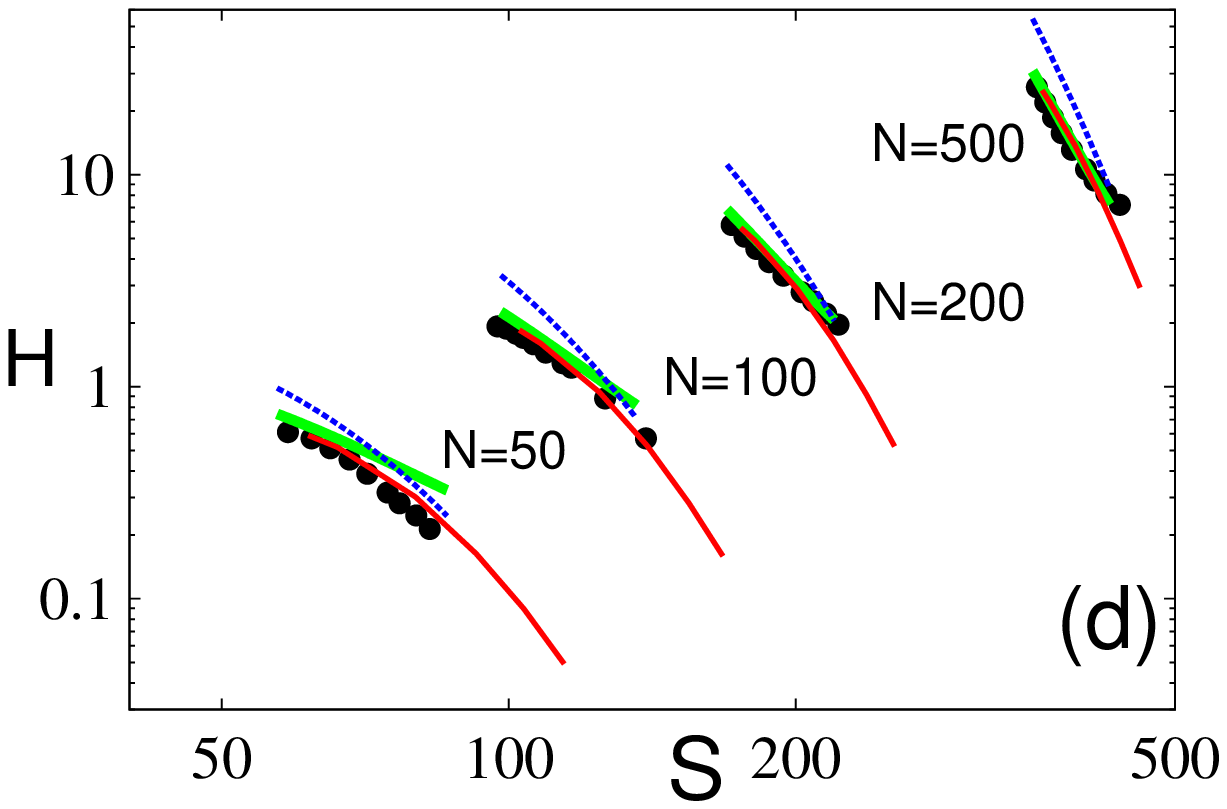}
\includegraphics[scale=0.5,angle=0]{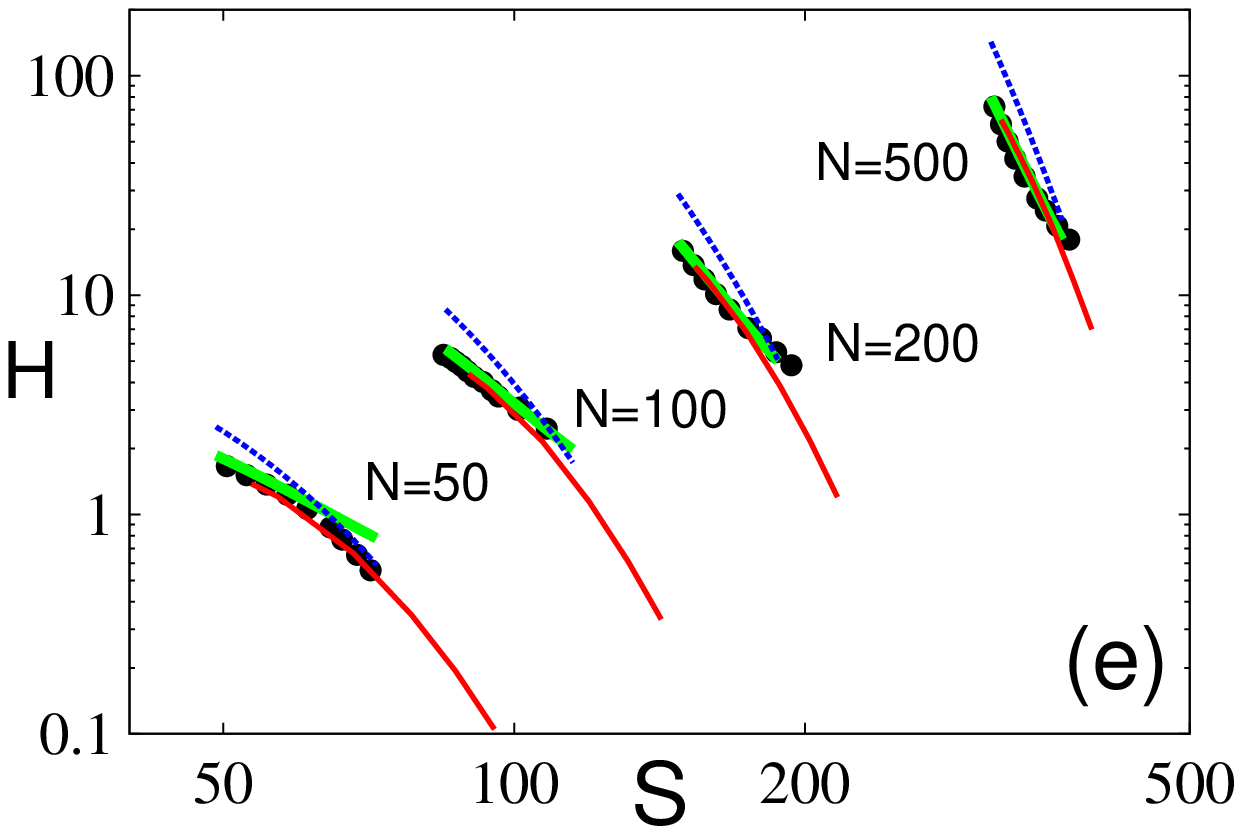}
\includegraphics[scale=0.5,angle=0]{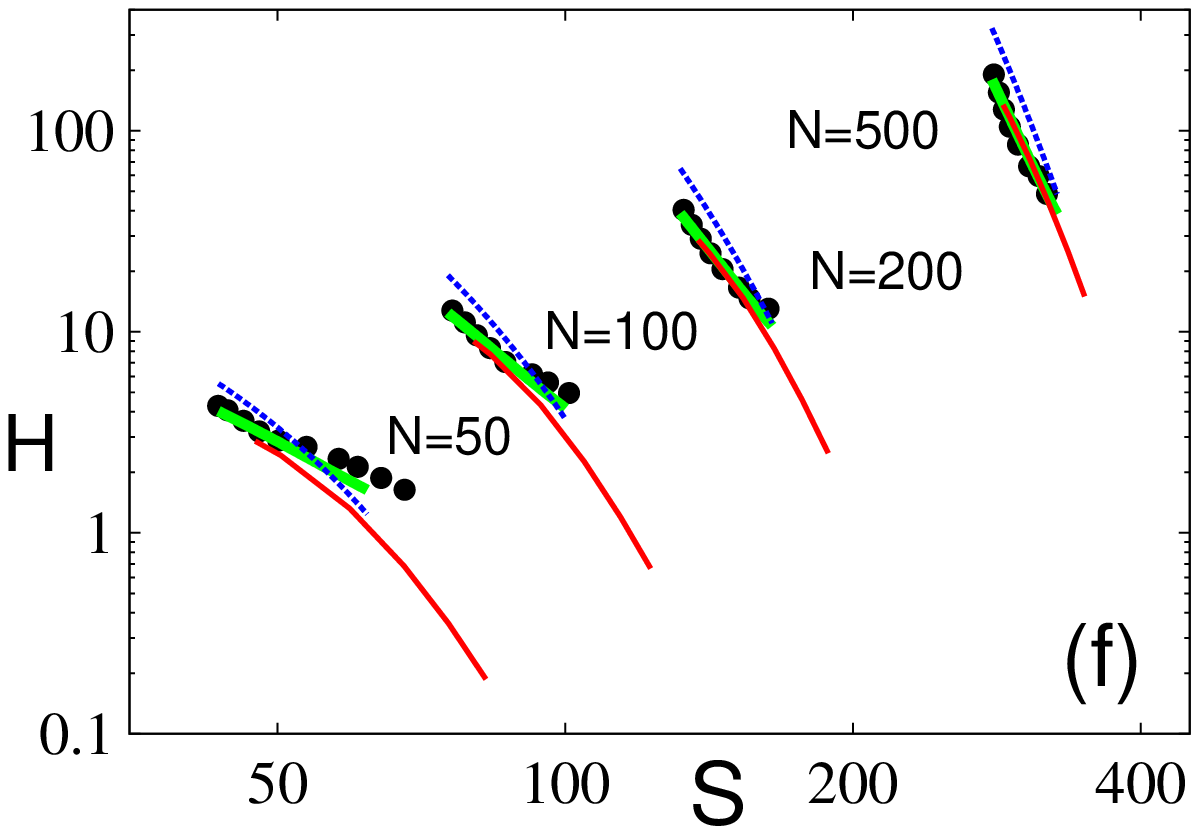}
\caption{\label{fig1}
Rf heating rates of nonneutral plasmas 
in a Paul trap as a function of cloud size $S$ 
for $N=50$, 100, 200, and 500 particles and 
(a) $q=0.10$, (b) $q=0.15$, (c) $q=0.20$, (d) $q=0.25$, 
(e) $q=0.30$, (f) $q=0.35$. Plot symbols: 
Molecular-dynamics simulations. Heavy green line: 
Analytical, temperature-dependent rf-heating formula. 
Dashed blue line: Analytical 
rf heating formula with equi-partition approximation. 
Thin red line: Mean-field approximation. 
}
\end{figure}
%
In this section we present our 
numerical 
and analytical 
methods for evaluating 
the heating rate 
$H=H(S,N,q)$. 
The numerical results obtained in 
Section~\ref{MDS} provide 
the target data sets to be matched by our analytical 
formulas developed in 
Section~\ref{AHF}. 
This succeeds to 
an excellent degree of accuracy. 
In Section~\ref{MFC}, we compare 
the results of our molecular-dynamics 
simulations with the solutions of a 
nonlinear mean-field equation. 
Excellent agreement between 
the rf heating rates obtained by these two 
qualitatively different numerical methods is obtained. 
This provides an independent check 
of our molecular-dynamics simulations. 
 
\subsection{Molecular dynamics simulations}
\label{MDS}
Using a 5th order Runge-Kutta method \cite{NUMREC}, we 
performed extensive mole\-cu\-lar-dynamics simulations 
\cite{MD} 
of (\ref{THEO1}), extracting rf heating rates 
as discussed in \cite{Tarnas} 
directly via computation of $dE/dt$ according to 
(\ref{THEO2}) 
for $N=50$, 100, 200, and 500 particles 
and $q=0.10$, 0.15, 0.20, 0.25, 0.30, and 
0.35. The resulting rf heating rates are shown 
as a function of cloud size $S$ 
as the black data points in Fig.~\ref{fig1}. 
While rf heating rates for $q=0.2$ were already 
computed and presented in \cite{Tarnas}, 
the data set displayed in Fig.~\ref{fig1}, 
covering six different $q$ values, is 
more extensive than 
the data set presented in \cite{Tarnas}. 
We checked that 
the rf heating rates in 
the $q=0.20$ panel of Fig.~\ref{fig1} are consistent 
with the rf heating rates presented in \cite{Tarnas}. 
 
\subsection{Analytical heating formulas} 
\label{AHF}
Defining the temperature 
\begin{equation}
T = \frac{1}{3} 
\lim_{n\rightarrow\infty} \frac{1}{n} 
\sum_{k=1}^n
\sum_{i=1}^N \dot{\vec R}_i^2(k\pi), 
\label{THEO4}
\end{equation}
of a plasma cloud in the stationary state \cite{Jones}, 
the kinetic energy of the plasma cloud may be 
evaluated immediately on the basis of (\ref{THEO1a}). 
The result is \cite{Jones} 
\begin{equation}
E_{\rm kin} = \frac{3N}{2}
\left(1+\frac{q^2}{4}\right) T + 
\frac{1}{2} q^2 S^2 ,  
\label{THEO5}
\end{equation}
where we assumed that 
the micro- and macromotions are uncorrelated. 
Since, in the stationary state, 
$\gamma$ determines 
the cloud size $S$, we may write $\gamma=\gamma(S,N,q)$. 
This way, if $\gamma(S,N,q)$ is known, 
we may use (\ref{THEO2}) and (\ref{THEO5}) to 
compute $H(S,N,q)$ analytically. 
 
Our analytical formula for $\gamma(S,N,q)$ is 
based on the $q$ and $N$ scaling of the 
critical gamma, $\gamma_c(N,q)$, at which 
the transition to the crystal occurs \cite{Danny1}. 
In \cite{Danny1} we found that for a given $q$, 
$\gamma_c$ scales like an iterated-log law in $N$. 
In order to 
reveal the $q$ dependence of $\gamma_c$, we performed 
extensive additional molecular-dynamics 
simulations and established that 
\begin{equation}
\gamma_c(N,q) = C(q)\ln[\ln(N)]-D(q),
\label{THEO6}
\end{equation}
where 
\begin{equation} 
C(q) = 1.31\times q^{4.57},\ \ \ D(q)=0.13\times q^{3.77} . 
\label{THEO7}
\end{equation}
Additional molecular-dynamics simulations 
then allowed us to determine the complete scaling 
of $\gamma(S,N,q)$. We found 
\begin{equation}
\gamma(S,N,q) = \gamma_c(N,q) 
\exp\left[ -2 q^{0.9} \left(\frac{S-S_c(N,q)}
{\sqrt{S_c(N,q)}}\right) \right], 
\label{THEO8}
\end{equation}
where $S_c(N,q)$ is the critical cloud size 
(the smallest possible cloud size) at $\gamma=\gamma_c$. 
Since $S_c$ is very close to the size
$S_{\rm crystal}(N,q)$ of the crystal, 
we write 
\begin{equation}
S_c(N,q) = S_{\rm crystal}(N,q) + \sigma(N,q), 
\label{THEO9}
\end{equation}
where \cite{Jones} 
\begin{equation}
S_{\rm crystal}(N,q) = \left(\frac{3}{5q^{4/3}}\right)^{1/2} 
N^{5/6}. 
\label{THEO10}
\end{equation}
For the shift function $\sigma$ we found empirically that 
\begin{equation}
\sigma(N,q) = [37.8 \ln(N) - 96.6] 
\exp\left\{ -[1.4\ln(N) + 3.1] q\right\}. 
\label{THEO11}
\end{equation}
On the basis of the above formulas, $\gamma(S,N,q)$ 
can now be assembled and expressed in closed form 
as an analytical function. 
 
The only missing ingredient 
for the use of (\ref{THEO5}) for the construction of 
an analytical representation of $H(S,N,q)$ is now a 
knowledge of $T(S,N,q)$. Based on 
our extensive numerical data sets we found 
\begin{equation}
T(S,N,q) = 0.7
\left[ \frac{\gamma(S,N,q)}{\gamma_c(N,q)} \right]^{-0.9},  
\label{THEO12}
\end{equation}
which is valid for temperatures $0.7\lesssim T\lesssim 3$. 
Since, at this point, both $\gamma(S,N,q)$ and 
$\gamma_c(N,q)$ are 
known analytically, $T(S,N,q)$ is known analytically, 
which implies that $H(S,N,q)$ is now known analytically. 
We may also use (\ref{THEO12}) together with 
(\ref{THEO8}) to obtain $S$ as a function of $T$, 
$N$, and $q$ according to 
\begin{equation}
S(T,N,q) = S_c(N,q) + 
\frac{\sqrt{S_c(N,q)}}{1.8\times q^{0.9}} 
\ln\left(\frac{T}{0.7}\right) .
\label{THEO12a}
\end{equation}

The heavy, green, solid lines in Fig.~\ref{fig1} 
represent $H(S,N,q)$ evaluated 
analytically in the range from $S_c(N,q)$ to 
$S(T=5,N,q)$ 
according to the formulas 
stated explicitly above. 
We see that for small $S$ (low temperature $T$) the 
analytical formula for $H(S,N,q)$ fits the numerical 
simulation data very well. 
The analytical lines start deviating from the numerical data 
for larger $S$ values, which we established to correspond 
to a temperature of $T\approx 3$. 
The deviation is due to our temperature formula 
(\ref{THEO12}), which is valid only for low temperatures 
close to the temperature where the cloud $\rightarrow$ crystal 
phase transition occurs \cite{Danny1}. 
Nevertheless, as seen in Fig.~\ref{fig1}, the temperature 
range up to $T=3$ covers a considerable range in $S$. 
 
For $T>3$ a different approach turned out to be useful. 
Assuming equipartition between the micromotion and 
the kinetic energy of the macromotion \cite{BaWa} 
turns (\ref{THEO5}) into 
\begin{equation}
E_{\rm kin} = q^2 S^2 .  
\label{THEO13}
\end{equation}
Using this expression for $E_{\rm kin}$ in (\ref{THEO2}) 
results in the dotted blue lines in Fig.~\ref{fig1}. 
We see that for small temperatures 
the assumption (\ref{THEO13}) of equi-partition is 
not as good as using the temperature-dependent 
formula (\ref{THEO5}), but improves markedly 
for $T>3$. 
Therefore, we recommend to compute $H(S,N,q)$ 
on the basis of (\ref{THEO5}) for $T<3$ and 
on the basis of (\ref{THEO13}) for $T>3$. 
While, as documented in Fig.~\ref{fig1}, 
this piecewise definition of our analytical 
formula for $H(S,N,q)$ works very well, we 
are currently working on an improved formula 
for $T(S,N,q)$, which will uniformly cover the 
entire range of $S$, $N$, and $q$ values. 
 
 
\subsection{Mean-field calculations}
\label{MFC} 
In order to check our numerical simulations, we 
solved the mean-field equation (55) in 
\cite{Jones} for $T=0.7$, 1, 2, 3, 4, and 5,  
for each of the $q$-$N$ 
combinations 
shown in Fig.~\ref{fig1}. The resulting 
heating rates are shown as the thin, red lines 
in Fig.~\ref{fig1}. 
The excellent agreement between the 
mean-field calculations and our numerical 
molecular-dynamics simulations 
in the temperature range $0.7\le T < 5$
shows that 
(i) our molecular-dynamics simulations are 
reliable and (ii) detailed and expensive 
molecular-dynamics simulations in this 
temperature regime are not 
necessary;  
a much cheaper mean-field calculation 
suffices. Although our analytical heating 
formulas (solid, green and dashed, 
blue lines in Fig.~\ref{fig1}) 
were computed over the same temperature range 
as our mean-field heating rates, 
Fig.~\ref{fig1} shows that the 
lines corresponding to our analytical results 
systematically terminate at smaller 
$S$ values than the mean-field heating rates 
(thin, red lines in Fig.~\ref{fig1}). 
The reason is the following. For computing the end points 
$S(T=5,N,q)$ of the lines that represent 
our analytical results, we used the analytical 
formula (\ref{THEO12a}). This formula, 
however, is based on our 
temperature formula (\ref{THEO12}), which, 
as mentioned above, 
is valid only for $T\lesssim 3$ and deteriorates 
for $T\gtrsim 3$. This results in a prediction of  
$S(T=5,N,q)$, which, typically, is of the order of 
20\% smaller than the $S$ value at $T=5$ predicted
by our mean-field calculations. The log scale 
used in Fig.~\ref{fig1} exaggerates this relatively 
small difference. 

\section{Discussion}
\label{DISC} 
To our knowledge, 
our analytical rf heating formulas 
for $H(S,N,q)$ are the first 
such formulas that comprehensively cover the 
entire parameter range of rf-driven nonneutral 
plasmas stored in a hyperbolic Paul trap. 
This is significant, since $H(S,N,q)$, over 
the $S$, $N$, and $q$ ranges shown in 
Fig.~\ref{fig1}, covers approximately 5 orders 
of magnitude. 

We are well aware of the fact that our analytical 
heating formula is not derived from first principles. 
This would be an exceptionally difficult task to 
accomplish, which, to date, has not even succeeded in 
the case $N=2$. Yet, our analytical heating formula 
is much more than, say, a fit of a multi-variable 
polynomial to the results of our heating simulations. 
The difference is that the fit functions we use 
are not arbitrary, but carefully extracted from 
the $S$, $N$, and $q$ dependence exhibited 
by the heating rates provided by our 
molecular-dynamics simulations. 
Thus, only the numerical constants are fitted, 
while the shape of the fit functions is dictated 
by our data. 
Therefore, although our heating formula 
is based on input information of 
the rf heating of nonneutral plasmas for only up to 
$N=500$ particles, but 
since it is based on scaling properties extracted 
directly from the data, our analytical heating formula 
has predictive power and remains valid for 
$N\gg 500$. We spot-checked this explicitly 
by comparing rf heating rates computed via both 
molecular-dynamics simulations and our 
analytical rf heating formula for $q=0.2$ 
and $N=1,000$, $N=2,000$, and 
$N=5,000$ particles, which, currently, 
is the limit of our computer resources. 
However, using our vastly 
faster mean-field calculations, allowed us 
to check the validity of our analytical rf 
heating formula for several values of 
$q$ and particle numbers up to 
$N=10^5$.  
 
Currently we are unable to extend our molecular-dynamics 
simulations with confidence beyond the 
$S$ values shown in Fig.~\ref{fig1}. This has 
two reasons. (i) According to (\ref{THEO8}) 
the damping constant required to 
establish a stationary state for large $S$
is exponentially small, requiring 
exponentially long simulation times, which 
are currently beyond our computer budget. 
(ii) For exponentially small $\gamma$,
the damping term in (\ref{THEO1}) is 
exponentially 
small compared with the other terms in 
(\ref{THEO1}) and is drowned out by 
numerical round-off noise. Keeping our 
current numerical methods, obtaining 
reliable numerical results for very small $\gamma$ 
can only be 
achieved by running our simulations in 
quadruple precision, which will extend 
the already exponentially long simulation 
times as discussed in (i). Nevertheless, 
exploring the large-$S$, large-$T$ regime 
is definitely on our agenda. 
 
While the small-$S$ regime, explored in 
this Letter, allows for a relatively simple, 
unified treatment in terms of particle dynamics 
and heating rates, the large-$S$ regime, 
currently inaccessible to our 
numerical simulation methods, may hold 
surprises. For increasing $T$, the plasma 
becomes increasingly more dilute; a tenuous, 
one-component, nonneutral plasma results, in 
which single-particle properties may dominate 
collective plasma properties. The result is 
a much richer, nonlinear dynamics, which may 
be impossible to capture with a single, 
analytical heating formula $H(S,N,q)$ with 
a simple closed-form analytical structure as 
presented in this Letter in the small-$S$, 
low-temperature regime. 
 
In this Letter we focused our discussion 
on the energy flow from the rf field to 
the trapped nonneutral plasma in the 
stationary state, established in the 
presence of a damping mechanism, 
modelled in (\ref{THEO2}) in terms of 
a damping term with damping constant $\gamma$. 
This is a situation we called situation (A) 
in Section~\ref{INTRO}. In this case, 
in addition to providing power to heat 
up the trapped plasma, 
the rf field has to provide power to 
sustain the micromotion in the presence of damping.  
The power necessary to sustain the micromotion is 
$\gamma q^2 S^2/2$. In situation (B), 
no damping is present, and $\gamma$ is zero. 
In this situation all the power provided 
by the rf field is used to heat up the plasma; 
providing power for sustaining the micromotion 
is not necessary in situation (B). 
Therefore, the plasma heating rate  
$H_P(S,N,q)$ may be obtained from the 
heating rate $H(S,N,q)$, pertaining to 
situation (A), via 
\begin{equation}
H_P(S,N,q) = H(S,N,q) - 
\frac{1}{2}\gamma(S,N,q) q^2 S^2, 
\label{DISC1}
\end{equation}
where $\gamma(S,N,q)$ is the damping constant 
necessary to achieve the stationary state 
with cloud size $S$. 
We verified the validity of (\ref{DISC1}) 
explicitly via molecular-dynamics simulations 
in which we first established the stationary 
state with the help of $\gamma(S,N,q)$, subsequently 
taking $\gamma$ to zero in (\ref{THEO1}), and then evaluating 
$H_P(S,N,q)$ during the $\gamma=0$ expansion phase of the cloud. 
 
In this Letter we restricted ourselves to the discussion 
of spherical plasma clouds, i.e., $a=q^2/2$. 
We are currently working on 
extending our results to the case of non-spherical 
plasma clouds in the hyperbolic Paul trap, and to the case of 
plasma clouds in 
the linear Paul trap \cite{qc1,qc2,qc3}. 
No new analytical or 
numerical tools have to be developed since the conceptual 
framework established in this Letter is applicable to 
all rf-trap architectures. 
 
\section{Summary, conclusions, and outlook} 
\label{CONC} 
In this Letter we present, for the 
first time, a comprehensive, explicit, analytical 
heating formula 
that allows us to compute rf heating rates in the 
form $H(S,N,q)$ [$H_P(S,N,q)$, respectively]
for all trapped, spherical,  
nonneutral plasmas in a hyperbolic Paul trap 
in the low-temperature regime. Our formula, 
defined piecewise in two branches ($T<3$ and 
$T>3$), shows excellent agreement 
with detailed microscopic, time-dependent 
molecular-dynamics calculations and 
time-independent mean-field simulations 
conducted in the low-temperature regime ($T<5$).  
Our analytical and numerical methods 
define a general framework that 
may be applied 
to the construction of rf heating formulas 
for all rf-trap architectures and cloud shapes. 
 
\section{Acknowledgment} 
\label{ACKN} 
YSN acknowledges support from 
ARO MURI award W911NF-16-1-0349.

\section*{Appendix: Conversion to SI units} 
\label{CONV} 
Dimensionless quantities, as used in the main body of 
this paper, are the most convenient choice if 
the general, universal features of a given system are 
emphasized. 
However, if used for practical applications and 
laboratory experiments, SI units are more convenient. 
 
If the trapped nonneutral plasma consists of particles of charge 
$Q$ and mass $m$, the control parameters 
$a$ and $q$ are given by \cite{BKQW} 
\begin{equation}
a = \frac{8QU_0}{m\Omega^2(r_0^2+2z_0^2)}, \ \ \ 
q = \frac{4QV_0}{m\Omega^2(r_0^2+2z_0^2)},
\label{CONV1}
\end{equation}
where $r_0$ and $z_0$ are the distances of the Paul trap's 
ring electrode and end-cap electrodes from the 
center of the trap, respectively, $U_0$ and $V_0$ 
are the dc and ac voltages applied to the trap, 
respectively, $\Omega=2\pi f$ is the angular frequency 
of the trap, and $f$ is the ac frequency of the trap 
in Hz. With these quantities, the unit of time is 
\begin{equation}
t_0 = \frac{2}{\Omega}, 
\label{CONV2}
\end{equation}
the unit of length is 
\begin{equation}
l_0 =  \left( \frac{Q^2} 
{\pi \epsilon_0 m \Omega^2} \right)^{1/3}   , 
\label{CONV3}
\end{equation}
the unit of energy is 
\begin{equation}
E_0 =  \frac{m l_0^2}{t_0^2}  , 
\label{CONV4}
\end{equation}
the unit of temperature is 
\begin{equation}
T_0 =  \frac{E_0}{k_B}   , 
\label{CONV5}
\end{equation}
where $k_B$ is Boltzmann's constant, 
and the unit of heating rate is 
\begin{equation}
H_0 =  \frac{E_0}{t_0}=\frac{ml_0^2}{t_0^3} . 
\label{CONV6}
\end{equation}
Thus, the heating rate 
$H^{(SI)}$ in SI units is computed from the 
dimensionless heating rate $H$ according to 
\begin{equation}
H^{(SI)}(S,N,q) =  H_0\, H(S,N,q) , 
\label{CONV7}
\end{equation}
and the temperature $T^{(SI)}$ in SI units is computed from 
the dimensionless temperature $T$ according to 
\begin{equation}
T^{(SI)}(S,N,q) =  T_0\, T(S,N,q) . 
\label{CONV8}
\end{equation}
%

\section*{References}

\end{document}